%% file: hep.tex
\documentstyle[12pt,openbib]{article}
\begin{document}
\title{Scaling Law for Baryon Coupling to its Current and its possible
applications}
\author{Jishnu Dey $^{1}$\thanks{E-mail: deyjm@giascl01.vsnl.net.in, work
supported  in part by DST grant no. SP/S2/K04/92 Govt. of India, also
supported initially by FAPESP of Brasil, permanent address : 1/10 Prince
Golam Md.  Road, Calcutta India 700 026, },  Mira Dey $^2$\thanks{Work
supported in part by DST grant no. SP/S2/K04/92 Govt. of India, also
supported initially by CAPES of Brasil, permanent address : 1/10 Prince Golam
Md. Road, Calcutta India 700 026, }, T.  Frederico, $^{3}$  and Lauro Tomio
${^4}$\thanks{E-mail: tomio@axp.ift.unesp.br}}
\vskip .5 cm
\date{\today }
\maketitle
{\it Abstract}
\vskip .1 cm 
The baryon- coupling to its current ($\lambda_{B}$), in conventional QCD sum
rule calculations (QCDSR), is shown to scale as the cubic power of the baryon
mass, $M_B$. Some theoretical justification for it comes from a simple
light-cone model and also general scaling arguments for QCD. But more
importantly, taken as a phenomenological ansatz for the present, this may
find very good use in current explorations of possible applications of QCDSR
to baryon physics both at temperature $T = 0$, $T \ne 0$ and/or density $\rho
= 0$, $\rho \ne 0$.

\vskip 1cm
(1) {\it International Centre for Theoretical Physics, Trieste, Italy 34100
and Azad Physics Centre, Dept. of Physics, Maulana Azad College, Calcutta,
India 700 013}  
\vskip .1 cm
(2) {\it International Centre for Theoretical Physics, Trieste, Italy 34100
and  Department of Physics, Presidency College, Calcutta, India 700 073}
\vskip .1 cm
(3) {\it Departamento de F\'\i sica, ITA, Centro T\'ecnico Aeroespacial,
12.228-900 S\~ao Jos\'e dos Campos, S\~ao Paulo, Brasil}
\vskip .1 cm
(4) {\it Instituto de F\'\i sica Te\'orica, Universidade Estadual Paulista, 
01405-900 S\~ao Paulo, S\~ao Paulo, Brasil}
\vskip .1 cm
PACS{24.85.+p, 12.39.Ki, 14.20.-c, 13.40.Gp, 11.55.Hx}
\newpage
The QCD sum rule calculation of Shifman et al. \cite{SVZ} reproduces some of
the predictions of QCD obtained in terms of fundamental degrees of freedom.
For reviews on the subject see Refs. \cite{shif},\cite{rry}, \cite{dd}.  One
of the outputs of QCD sum rule calculations \cite{ioff81} is the baryon
coupling to its quark content. For example, denoting the nucleon as N, the
current constructed out of quarks
\begin{equation}
{\cal J}_N(x) = (\overline  u^{a}(x)C\gamma_{\mu}u^b(x))
\gamma_5\gamma_{\mu}d^c(x)\varepsilon^{abc}
\end{equation}
carries the quantum numbers of the nucleon, where $C$ is the charge
conjugation operator, $a,b,c$ are colour indices, $\varepsilon$ is the
antisymmetric tensor and $u(x)$ and $d(x)$ are the up and down quark
operators. The matrix element 
\begin{equation}
\langle 0 |{\cal J}_N| N \rangle = \lambda _N \; v^{(r)}
\end{equation}
where $ \lambda _N$ is the coupling we are concerned with and $v^{(r)}$ is the
usual Dirac spinor for polarization $r$, normalized to the nucleon mass $M_N$ as :
\begin{equation}
{\bar v^{(r)}}v^{(r)} = 2 M_N.
\end{equation}
Similarly one can define a coupling $\lambda _B$ for each baryon $B$ with a
mass $M _B$. We find 
\begin{equation}
\lambda _B \; = constant \; M_B^{3}
\label{eq:m3}
\end{equation}
from the existing determination of the sum rules as shown in Fig. 1 which is
discussed below.

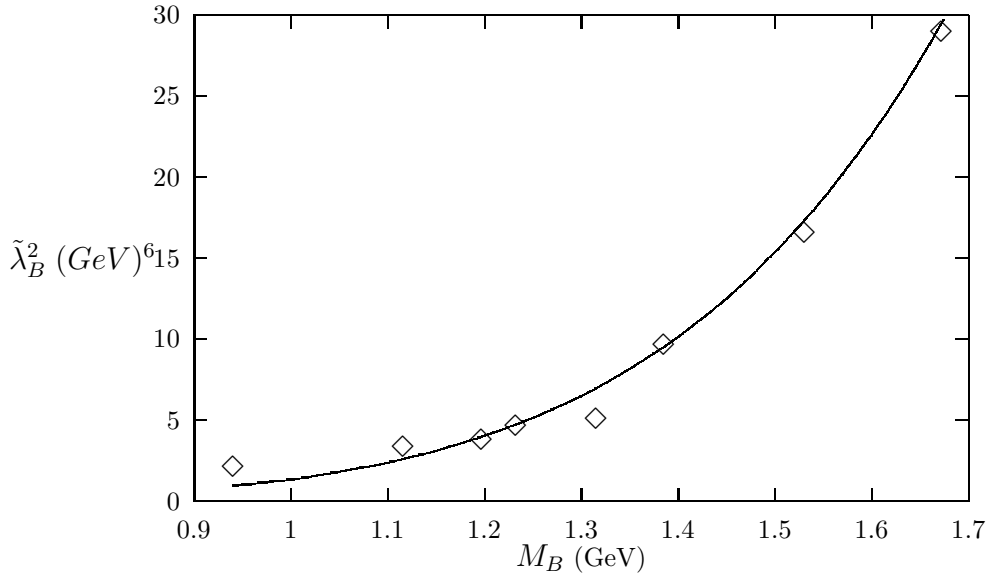
\begin{figure}
\begin{center}
\input{st.tex}
\end{center}
\caption{ The points are extracted from QCDSR calculations and the line is our model.}
\end{figure}

Besides $N$, the couplings of $\Lambda $, $\Sigma $, $\Xi $  given
by Ioffe and Smilga \cite{smil} and $\Omega $ by Nielsen et al. \cite{mn} to
their currents are known.  Using calculations of Reinders et al. \cite{re}
(also reported in \cite{rry} and \cite{shif}) one can extract the couplings
$\Sigma^*$ and $\Xi^*$. We take $\Delta$ from Belyaev and Ioffe \cite{bi}.
Fig. 1 shows that they present a scaling law with the baryon mass, that cover
a wide range of the spectrum and are practically independent of the spin
structure of the baryon in so far as we find it to be valid for S = 1/2 octet
as well as for S = 3/2 decuplet members.

From \cite{smil}, $\tilde \lambda^2 _B \equiv 2(2\pi)^4 \lambda _B^2 $, are
given respectively for $N$, $\Lambda $, $\Sigma $ and $\Xi $ as 2.1, 3.3, and
3.7 $GeV^6$.  For $\Omega $ the same quantity is 28.9 $GeV^6$ \cite{mn}.  For
$\Delta$, $\Sigma^*$ and $\Xi^*$ the respective values are 4.6, 9.6 and 16.5
$GeV^6$.  We find that the QCDSR results can be fitted very well to a simple
formula
\begin{equation}
\tilde \lambda^2_B  = (1.16)^2 \; M_{B}^6. 
\label{eq:m6}
\end{equation}
To determine $ \tilde \lambda^2 _B$ is not easy in QCDSR, as the reader can
find in the literature \cite{ioff81,smil,bi}. The final value given by
\cite{smil} gives a slight overestimate of both the proton and the neutron
magnetic moments.  The nucleon results that we report are for the choice of a
current due to Ioffe \cite{ioff81} (reprinted and supported in \cite{shif}).

It is clear for us that the scaling law we have found follows from
general qualitative considerations which can be at various levels of
sophistication. Thus for example we can argue that:

(1) in a simple quantum mechanical sense the quark is normalized in a volume
proportional to the Compton wavelength $\sim 1/M$, where M is the mass of the
constituent quark which on the average is proportional to the baryon mass,
$M_B$. The probability of finding 3 such particles is the cube of this
factor. This leads to the eq. (\ref{eq:m3}).

(2) in a more sophisticated sense the light quark sector of QCD is
conformally invariant with the invariance broken only by the scale of the
constituent quark mass ($\approx 1/3 M_B$) so that this mass controls the
scaling of all relevant quantities.

The coupling of baryons to their currents are crucial for  all experimental
properties obtained with QCDSR calculations. They are involved in deriving
observables like magnetic moments and electromagnetic and weak transitions,
etc.  The consequences of the scaling can be verified by a systematic
calculation of hadronic properties.

The knowledge of some systematics of  how the coupling $\lambda _B$
changes from one baryon to another could have further use. To understand this
we must recall the basic structure of a fermion sum rule, namely that it
always comes in a pair. This is due to the Dirac structure of the propagator
which has a scalar part and another propotional to the Dirac  $\gamma$ matrix.
They are called the chiral odd and the chiral even sumrules \cite{jin1}. The
nucleon mass was originally extracted by Ioffe \cite{ioff81} with a combined
use of both the sum rules. It is necessary to use these two sum rules
together since there are two unknowns, $\lambda _B$ and $M_B$. 

The QCDSR depend on the parameter $M$, called the Borel parameter which is
very useful for the phenomenological side of the sum rule where it cuts off
high energy contributions with a factor $exp(-s/M^2)$, where $s$ is the
squared momentum transfer. On the QCD side also it leads to a function of
$M^2$. The sum rule should be plotted as a function of $M$ and a `window' is
to be found where there is stability with variation of this parameter. It was
soon clear, (see for example \cite{jin1}) that one of the sumrules work well
while the other fails, in case of baryons. This pattern is also seen in  the
sum rules for the matrix elements of electromagnetic current and axial vector
current (for references see \cite{jin1}). The most important contribution in
any given channel is the ground state baryon. However the excited states also
contribute in higher order. The different behaviour of various sumrules can
be traced to the fact that even and odd parity states contribute with
different signs.  If chiral symmetry is realized in the Wigner-Weyl mode at
high energies, i.e.  by parity doubling, it is possible to have either
cancellation or reinforcement between excited state configurations.
Irrespective of the exact manner in which the cancellations  take place it is
clear that the odd sumrule favours the extraction of the baryon ground state
mass. The point has been reemphasized in a recent paper by Jin and Tang
\cite{jin2}.

Belyaev and Ioffe \cite{bi}, \cite{bel} extended the nucleon sumrule
calculation to determine the mass splitting between hyperons (Y) and nucleons
treating the strange quark mass as a perturbation.  In addition to the mass
shift $M_Y - M_N$, one must also take into account the change in the coupling
constant $\lambda _Y^2 -\lambda _N^2$ and the change in the continuum
threshold, $s_0$. It is found \cite{smilga} that $s_0$ goes according to the
`mnemonic rule' :
\begin{equation}
s_0 = ( M_B\; +\; 0.5\; GeV)^2.
\end{equation}
Therefore with our scaling law it is possible to extract masses from one
set of sum rules only. The transitions and magnetic moments can also be
better determined using the odd chiral sum rules, given the value of the
coupling from the scaling law. 

The above results were found while looking at a simple relativistic Faddeev
model for the nucleon, which therefore serves as some sort of a justification
for it also. The model uses a null-plane ($x^+=ct+z=0$)  wave function for
the baryon and treats some boosts in a consistent way.   The confining models
like the MIT bag or the soliton bag models have the complication of requiring
elaborate center of mass projection to restore covariance, a feature usually
shared by the one-body confining potentials used in the literature.  To
obtain baryon form-factors at high momentum transfers, one should use a
consistent boosted wave function. Relativistic constituent quark models, with
null-plane wave functions \cite{teren}, offer such an opportunity.  They are
covariant under kinematical front-form boosts
\cite{keispol}.  This point is in fact a deeper consequence of the stability
of the Fock-state decomposition of the wave function under such
transformations \cite{wil90}.

The three quark null-plane bound state is obtained through a renormalized
effective zero-range attractive force \cite{pach95}.  Being bound to the
scale of the hadron size, the constituent quarks have well defined masses. As
a consequence of the contact interaction, only the momentum scales are important
and it allows to identify the minimum dynamical inputs in the constituent
quark model. The advantage is that the strength of the contact interaction is
scaled out of the calculation and thus no extra scale parameters like the bag
constant have to be used.

The constituent quark in the null -plane, $|Q\rangle $, can be described as a
sum over current quark and current antiquark ($|q\rangle $, $ |\bar q\rangle$)
Fock-states:
\begin{equation}
|Q\rangle  = A_q|q\rangle  + A_{qq\overline q}|qq\overline q\rangle  + ..... 
\label{fock}
\end{equation}
and the integration of $|A_q|^2$ over the current quark phase-space gives
the probability, $\eta$, of a constituent quark to be the current quark. 
This $\eta$ is probed in the pion weak decay constant and  pion
deep-inelastic structure function \cite{fm}.

Considering that 
the broad scaling law is not dependent on the details of spin or isospin 
structure of the baryon, and to focus on the gross scaling, we assumed that
the spin and isospin have no 
dynamical effect. \ The total antisymmetrization can be achieved by
antisymmetrizing the color-isospin-spin degree of freedom, leading to a 
totally symmetric spatial wave function.
The momentum canonically associated with the
position coordinates of the $i$-th quark, in the  null-plane, are the
kinematical momenta, $k_i^+ $ and $\vec  k_{i\perp} $. The corresponding
momentum fraction is defined as $x_i \ = \ k_i^+/P^+$, where $P^+$ is the
$(+)$ component of the total momentum of the baryon. The momentum
constraints, in the center of mass of the three quarks, are given by
\begin{equation} 
x_1 +x_2 + x_3 \ = \ 1  \> \> \mathop{and} \> \> 
\vec  k_{1 \perp} + \vec  k_{2\perp} + \vec  k_{3\perp} \ = \ 0
\ .
\label{xk}\end{equation}
With the above variables we obtain the three quark bound-state null-plane 
wave function in terms of Faddeev components of the vertex, 
$ v(x, \vec  k_\perp) $\cite{pach95}:
\begin{eqnarray}\Psi (x_1, \vec  k_{1 \perp}; x_2, \vec 
k_{2\perp})= \frac{ v (x_1, \vec  k_{1\perp}) + v (x_2, \vec 
k_{2\perp})+ v (x_3, \vec  k_{3\perp})}{\sqrt{x_1 x_2 x_3} \left[
M^2_B \ - \ \sum_{j=1,3} (k^2_{j\perp} + M^2)/x_j \right] },
\label{1}\end{eqnarray} 
where $M_B$ is the baryon mass and $M$ is the constituent quark mass.
The normalization is chosen such that
\begin{eqnarray}
\int dx_1dx_2d^2k_{1\perp}d^2k_{2\perp} 
\left[\Psi (x_1, \vec  k_{1 \perp}; x_2, \vec k_{2\perp})\right]^2
= 1 .\label{normal}
\end{eqnarray} 

The vertex component, $v$, satisfies a Weinberg-type \cite{wein} integral
equation \cite{pach95,tob92}, in which the subsystem scattering is summed up
to all orders in the ladder approximation, following the original Faddeev
construction of the three-body connected kernel equations \cite{fadd}. 
In the baryon center of mass system, the vertex $v(y,\vec  p_\perp )$
is given by 
\begin{eqnarray}
v(y,\vec  p_\perp)= \frac{i}{(2\pi)^3}
\tau(M_2)\int^{1-y}_\frac{M^2}{M^2_B}\frac{dx}
{x(1-y-x)}\int^{k_\perp^{max}}d^2k_\perp\frac{v(x,\vec 
k_\perp)}{M^2_B-M^2_3} \ ,\label{4}
\end{eqnarray} 
where the kinematical momentum $x$, $\vec  k_\perp $ and $y$, 
$\vec  p_\perp$ describe the spectator quark states in the initial and 
final vertex, respectively. 
$M_2$ is the two-quark subsystem mass, 
\begin{eqnarray} 
M^2_2=(1-y)\left(M_B^2-\frac{p^2_\perp+M^2}{y}\right)
-p^2_\perp \ .
\label{5}
\end{eqnarray}
$M_3$ is the mass of the propagating virtual three-quark intermediate 
state, given by
\begin{eqnarray}M^2_{3}=\frac{k^2_\perp+M^2}{x}+
\frac{p^2_\perp+M^2}{y} + 
\frac{(\vec  p+ \vec  k)^2_\perp +M^2}{1-y-x} \ . 
\label{6}
\end{eqnarray}

The two-quark amplitude $\tau(M_2)$ for the contact interaction
needs to be renormalized (see Ref.\cite{renor}). 
In this process, the strength of the interaction
is fixed by the two-body (diquark) bound-state mass, 
which removes the infinities in  the physical two-body 
scattering amplitude \cite{tob92}.
The diquark bound-state mass $\mu$ is also equal to the $\overline Q \ Q $ 
- meson mass, because the pole of the two-quark scattering amplitude 
corresponds to a pole in the exchange channel. 

The expression for the two-quark amplitude, is given by\cite{tob92}: 
$$
\begin{array}{ll}
\tau(M_2) &  \nonumber \\ 

 =\;-i (2\pi)^2  \left[
\sqrt{\frac{M^2}{\mu^2}- 
\frac{1}{4}} \arctan \left(\frac{\mu}
{\sqrt{4M^2-\mu^2}}\right) 
-\sqrt{\frac{M^2}{M_2^2}- \frac{1}{4}} \arctan \left(\frac{M_2}
{\sqrt{4M^2-M_2^2}}\right) \right]^{-1}&

\end{array}
$$
where $M_2 < 2M$.

Considering that $ M^2_2 > 0$ in Eq.(\ref{5}), the spectator 
transverse momentum $k_\perp$ of Eq.(\ref{4}), attains a maximum value at 
$ k_\perp^{max}=\sqrt{(1-x)(M_B^2x-M^2)}$.

The integral equation, for the vertex in the new frame Eq.(\ref{4}), can be
rewritten by changing the corresponding variables.  This is possible because
of the nature of the kinematical transformations.  The mass of the
three-quark system, calculated from Eq.(\ref{4}), remains the same in any
frame connected to the center of mass frame by a kinematical transformation.

We define the relativistic coupling of the baryon wave function as the
integration in the phase-space, $x$ and $\vec k_\perp$, of  the Faddeev
vertex component:

\begin{eqnarray}\lambda_{LC} = 
M\int^{1-\frac{M^2}{M^2_B}}_\frac{M^2}{M^2_B}
\frac{dx}{x}\int^{k_\perp^{max}}d^2k_\perp v(x,\vec  k_\perp) \ .
\label{13}\end{eqnarray}

To compare with $\tilde \lambda^2_B$ (Eq.(\ref{eq:m6})) we revert back to
current quark picture.  Since each $|Q\rangle  \sim \sqrt{\eta} |q\rangle$ the
corresponding coupling in terms of current quarks is given by $\eta ^{3/2}
\lambda_{LC}$. Squaring and multiplying with the  factor $2(2\pi)^4$ we get
the corresponding quantity :
 
\begin{eqnarray}\tilde \lambda^2_{LC}= 2 (2\pi)^4 \eta ^3 \lambda^2_{LC}.
\label{19}\end{eqnarray} 

 There are only two parameters
in the model, namely the constituent quark and the diquark masses, $M$ and
$\mu$ respectively. They can be further reduced to one parameter, namely the
ratio, $\mu /M $ by calculating the radius. The electromagnetic nucleon
radius $r_p$ is obtained from the proton electric form factor $G_E(q^2)$, as
detailed in \cite{pach95}. $\mu /M = 1.9$ gives a good fit to $G_E(q^2)$
and corresponding ratio of the baryon mass to  quark mass is 2.58.

We remind the readers that the experimental value of the product of the proton
charge-radius ($r_p$) and the nucleon mass ($M_N$) is difficult to
reproduce, for example, in non-topological soliton models \cite{ban94}. 
It is nice to note that this model gives the product $r_p M_N (\sim 3.8)$ which is
near the experimental value.

We notice in Fig.2 that the minimum of $r_p M_N$ as a function of $\tilde
\lambda^2 _{LC}$, gives approximately the experimental value of $r_p M_N
(\sim 3.8)$.  Corresponding value of $\tilde \lambda ^2 _{LC}$ depends on
what value we take for $\eta $.  For example, it is $\sim \;1.46$ for $\eta=
1/2$, as shown in Fig. 2,  but for $\eta \sim .43 $ it is nearly unity.

It is satisfying to note that  $\eta \sim \frac{1}{2}$, is in agreement with
various numbers obtained in  \cite{fm} for the pion, under different choices
of constituent quark wave packet (two mass choices for Gaussian type 0.741
and 0.476, or same choices for hydrogen-like 0.463 and 0.408).

If we assume that the SU(3) flavour breaking does not change the ratio $\mu/M$
and the $\eta$ we obtain
$\tilde\lambda^2_{LC}$ as a function of the ground state baryon mass $M_B$.
By fixing $\eta = 0.43$ we get eq.(\ref{eq:m6}).

A glance at Fig. 1 shows that the decuplet baryons obey the scaling law more
consistently than the octet. This may be due to the fact that for the
decuplet baryons the instanton contribution is small \cite{shur} and
therefore the sum rule determination is free from the problems suffered by
the octet baryons.

Coming back to QCDSR, there is a recent calculation trying to incorporate
instanton effects into the nucleon sum rule, see Ref.\cite{dor}. According
to \cite{shif,dor}, the instanton effect is minimized in the Ioffe current.

In summary, we find evidence for a simple scaling law for the coupling of the
decuplet as well as the octet baryons to their current which scales with
the baryon masses with appropriate power fixed by its dimension. Some general
arguments and a simple relativistic Faddeev jutifies it.  Since $\tilde \lambda^2_B$ enters in calculations of magnetic
moments \cite{smil} and transitions \cite{mn} this scaling will in future be
used to obtain approximate relations between properties of the different
members of the 56-multiplet of baryons.

We acknowledge comments from Dr. A. Dorokhov. J. D. and M. D. wishes to
acknowledge the stimulating atmosphere of the International Centre for
Theoretical Physics, Trieste, Italy where this work was completed. The work
of T. F. and L. T. was partially supported  by the Conselho Nacional de
Desenvolvimento Cient\'{\i}fico e Tecnol\'ogico - CNPq  and Funda\c c\~ao de
Amparo \`a Pesquisa do Estado de S\~ao Paulo- FAPESP, Brasil.

\vskip 1cm
{\bf Figure captions :} 

Fig.1: The points are extracted from QCDSR calculations and the line is our
model. 

Fig.2: $r_p M_N$ as a function of $\tilde \lambda^2_{LC}$.
The constituent quark masses ($ M $) and meson masses ($ \mu $) have been
varied in the calculations, while the baryon mass was kept fixed at 0.938
GeV. The calculated points are shown with dots.
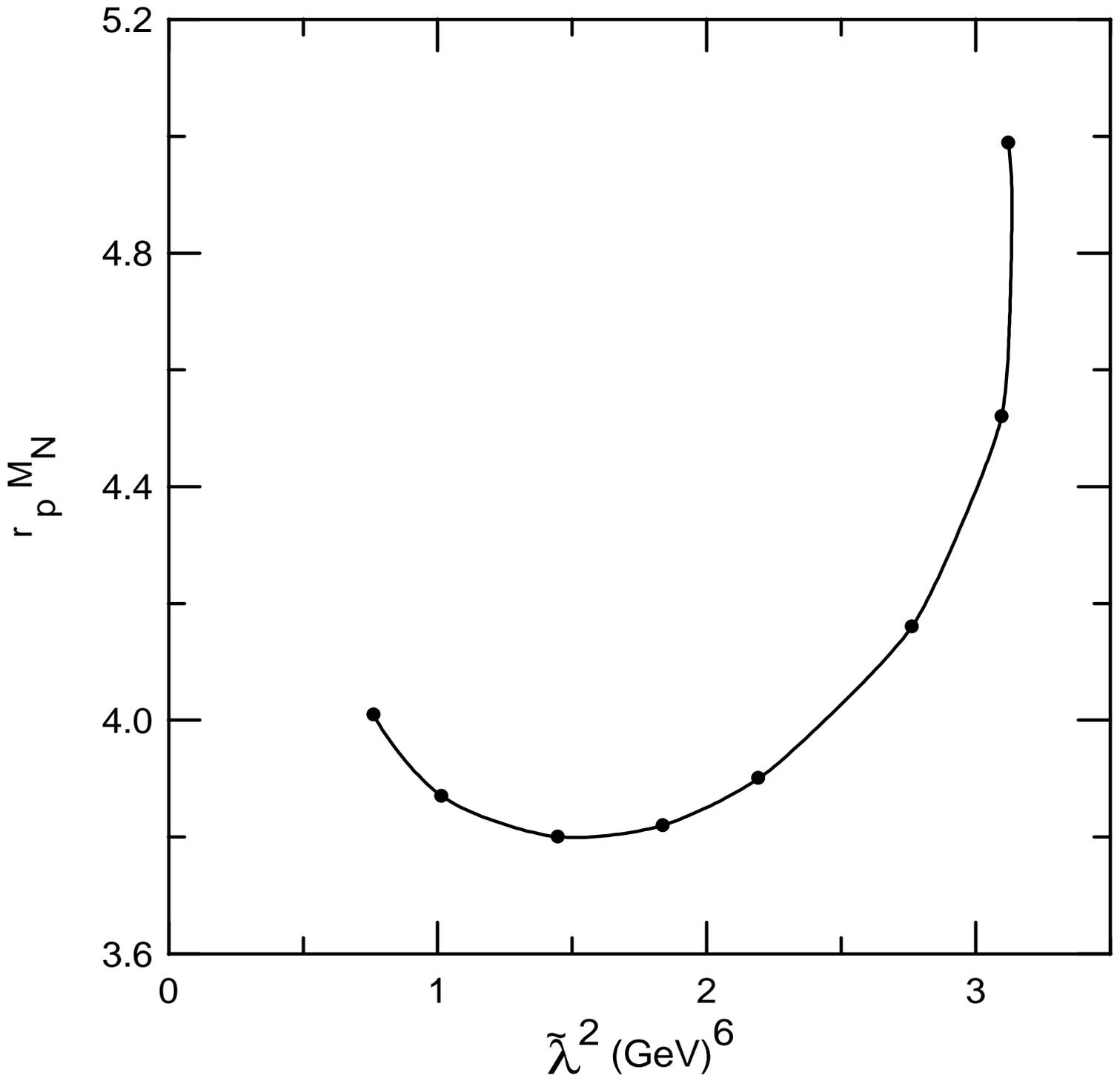
\end{document}

%% file: st.tex
\setlength{\unitlength}{0.240900pt}
\ifx\plotpoint\undefined\newsavebox{\plotpoint}\fi
\sbox{\plotpoint}{\rule[-0.200pt]{0.400pt}{0.400pt}}%
\begin{picture}(1500,900)(0,0)
\font\gnuplot=cmr10 at 10pt
\gnuplot
\sbox{\plotpoint}{\rule[-0.200pt]{0.400pt}{0.400pt}}%
\put(220.0,113.0){\rule[-0.200pt]{292.934pt}{0.400pt}}
\put(220.0,113.0){\rule[-0.200pt]{4.818pt}{0.400pt}}
\put(198,113){\makebox(0,0)[r]{0}}
\put(1416.0,113.0){\rule[-0.200pt]{4.818pt}{0.400pt}}
\put(220.0,240.0){\rule[-0.200pt]{4.818pt}{0.400pt}}
\put(198,240){\makebox(0,0)[r]{5}}
\put(1416.0,240.0){\rule[-0.200pt]{4.818pt}{0.400pt}}
\put(220.0,368.0){\rule[-0.200pt]{4.818pt}{0.400pt}}
\put(198,368){\makebox(0,0)[r]{10}}
\put(1416.0,368.0){\rule[-0.200pt]{4.818pt}{0.400pt}}
\put(220.0,495.0){\rule[-0.200pt]{4.818pt}{0.400pt}}
\put(198,495){\makebox(0,0)[r]{15}}
\put(1416.0,495.0){\rule[-0.200pt]{4.818pt}{0.400pt}}
\put(220.0,622.0){\rule[-0.200pt]{4.818pt}{0.400pt}}
\put(198,622){\makebox(0,0)[r]{20}}
\put(1416.0,622.0){\rule[-0.200pt]{4.818pt}{0.400pt}}
\put(220.0,750.0){\rule[-0.200pt]{4.818pt}{0.400pt}}
\put(198,750){\makebox(0,0)[r]{25}}
\put(1416.0,750.0){\rule[-0.200pt]{4.818pt}{0.400pt}}
\put(220.0,877.0){\rule[-0.200pt]{4.818pt}{0.400pt}}
\put(198,877){\makebox(0,0)[r]{30}}
\put(1416.0,877.0){\rule[-0.200pt]{4.818pt}{0.400pt}}
\put(220.0,113.0){\rule[-0.200pt]{0.400pt}{4.818pt}}
\put(220,68){\makebox(0,0){0.9}}
\put(220.0,857.0){\rule[-0.200pt]{0.400pt}{4.818pt}}
\put(372.0,113.0){\rule[-0.200pt]{0.400pt}{4.818pt}}
\put(372,68){\makebox(0,0){1}}
\put(372.0,857.0){\rule[-0.200pt]{0.400pt}{4.818pt}}
\put(524.0,113.0){\rule[-0.200pt]{0.400pt}{4.818pt}}
\put(524,68){\makebox(0,0){1.1}}
\put(524.0,857.0){\rule[-0.200pt]{0.400pt}{4.818pt}}
\put(676.0,113.0){\rule[-0.200pt]{0.400pt}{4.818pt}}
\put(676,68){\makebox(0,0){1.2}}
\put(676.0,857.0){\rule[-0.200pt]{0.400pt}{4.818pt}}
\put(828.0,113.0){\rule[-0.200pt]{0.400pt}{4.818pt}}
\put(828,68){\makebox(0,0){1.3}}
\put(828.0,857.0){\rule[-0.200pt]{0.400pt}{4.818pt}}
\put(980.0,113.0){\rule[-0.200pt]{0.400pt}{4.818pt}}
\put(980,68){\makebox(0,0){1.4}}
\put(980.0,857.0){\rule[-0.200pt]{0.400pt}{4.818pt}}
\put(1132.0,113.0){\rule[-0.200pt]{0.400pt}{4.818pt}}
\put(1132,68){\makebox(0,0){1.5}}
\put(1132.0,857.0){\rule[-0.200pt]{0.400pt}{4.818pt}}
\put(1284.0,113.0){\rule[-0.200pt]{0.400pt}{4.818pt}}
\put(1284,68){\makebox(0,0){1.6}}
\put(1284.0,857.0){\rule[-0.200pt]{0.400pt}{4.818pt}}
\put(1436.0,113.0){\rule[-0.200pt]{0.400pt}{4.818pt}}
\put(1436,68){\makebox(0,0){1.7}}
\put(1436.0,857.0){\rule[-0.200pt]{0.400pt}{4.818pt}}
\put(220.0,113.0){\rule[-0.200pt]{292.934pt}{0.400pt}}
\put(1436.0,113.0){\rule[-0.200pt]{0.400pt}{184.048pt}}
\put(220.0,877.0){\rule[-0.200pt]{292.934pt}{0.400pt}}
\put(45,495){\makebox(0,0){${\tilde \lambda}_B^2$ $(GeV)^6$}}
\put(828,23){\makebox(0,0){$M_B$ (GeV)}}
\put(220.0,113.0){\rule[-0.200pt]{0.400pt}{184.048pt}}
\put(281,137){\usebox{\plotpoint}}
\put(281,137.17){\rule{4.700pt}{0.400pt}}
\multiput(281.00,136.17)(13.245,2.000){2}{\rule{2.350pt}{0.400pt}}
\multiput(304.00,139.61)(4.704,0.447){3}{\rule{3.033pt}{0.108pt}}
\multiput(304.00,138.17)(15.704,3.000){2}{\rule{1.517pt}{0.400pt}}
\put(326,142.17){\rule{4.700pt}{0.400pt}}
\multiput(326.00,141.17)(13.245,2.000){2}{\rule{2.350pt}{0.400pt}}
\multiput(349.00,144.61)(4.927,0.447){3}{\rule{3.167pt}{0.108pt}}
\multiput(349.00,143.17)(16.427,3.000){2}{\rule{1.583pt}{0.400pt}}
\multiput(372.00,147.61)(4.927,0.447){3}{\rule{3.167pt}{0.108pt}}
\multiput(372.00,146.17)(16.427,3.000){2}{\rule{1.583pt}{0.400pt}}
\multiput(395.00,150.60)(3.259,0.468){5}{\rule{2.400pt}{0.113pt}}
\multiput(395.00,149.17)(18.019,4.000){2}{\rule{1.200pt}{0.400pt}}
\multiput(418.00,154.60)(3.113,0.468){5}{\rule{2.300pt}{0.113pt}}
\multiput(418.00,153.17)(17.226,4.000){2}{\rule{1.150pt}{0.400pt}}
\multiput(440.00,158.60)(3.259,0.468){5}{\rule{2.400pt}{0.113pt}}
\multiput(440.00,157.17)(18.019,4.000){2}{\rule{1.200pt}{0.400pt}}
\multiput(463.00,162.60)(3.259,0.468){5}{\rule{2.400pt}{0.113pt}}
\multiput(463.00,161.17)(18.019,4.000){2}{\rule{1.200pt}{0.400pt}}
\multiput(486.00,166.60)(3.259,0.468){5}{\rule{2.400pt}{0.113pt}}
\multiput(486.00,165.17)(18.019,4.000){2}{\rule{1.200pt}{0.400pt}}
\multiput(509.00,170.59)(2.491,0.477){7}{\rule{1.940pt}{0.115pt}}
\multiput(509.00,169.17)(18.973,5.000){2}{\rule{0.970pt}{0.400pt}}
\multiput(532.00,175.59)(1.937,0.482){9}{\rule{1.567pt}{0.116pt}}
\multiput(532.00,174.17)(18.748,6.000){2}{\rule{0.783pt}{0.400pt}}
\multiput(554.00,181.59)(2.491,0.477){7}{\rule{1.940pt}{0.115pt}}
\multiput(554.00,180.17)(18.973,5.000){2}{\rule{0.970pt}{0.400pt}}
\multiput(577.00,186.59)(2.027,0.482){9}{\rule{1.633pt}{0.116pt}}
\multiput(577.00,185.17)(19.610,6.000){2}{\rule{0.817pt}{0.400pt}}
\multiput(600.00,192.59)(1.713,0.485){11}{\rule{1.414pt}{0.117pt}}
\multiput(600.00,191.17)(20.065,7.000){2}{\rule{0.707pt}{0.400pt}}
\multiput(623.00,199.59)(1.713,0.485){11}{\rule{1.414pt}{0.117pt}}
\multiput(623.00,198.17)(20.065,7.000){2}{\rule{0.707pt}{0.400pt}}
\multiput(646.00,206.59)(1.637,0.485){11}{\rule{1.357pt}{0.117pt}}
\multiput(646.00,205.17)(19.183,7.000){2}{\rule{0.679pt}{0.400pt}}
\multiput(668.00,213.59)(1.484,0.488){13}{\rule{1.250pt}{0.117pt}}
\multiput(668.00,212.17)(20.406,8.000){2}{\rule{0.625pt}{0.400pt}}
\multiput(691.00,221.59)(1.484,0.488){13}{\rule{1.250pt}{0.117pt}}
\multiput(691.00,220.17)(20.406,8.000){2}{\rule{0.625pt}{0.400pt}}
\multiput(714.00,229.59)(1.310,0.489){15}{\rule{1.122pt}{0.118pt}}
\multiput(714.00,228.17)(20.671,9.000){2}{\rule{0.561pt}{0.400pt}}
\multiput(737.00,238.59)(1.310,0.489){15}{\rule{1.122pt}{0.118pt}}
\multiput(737.00,237.17)(20.671,9.000){2}{\rule{0.561pt}{0.400pt}}
\multiput(760.00,247.58)(1.121,0.491){17}{\rule{0.980pt}{0.118pt}}
\multiput(760.00,246.17)(19.966,10.000){2}{\rule{0.490pt}{0.400pt}}
\multiput(782.00,257.58)(1.173,0.491){17}{\rule{1.020pt}{0.118pt}}
\multiput(782.00,256.17)(20.883,10.000){2}{\rule{0.510pt}{0.400pt}}
\multiput(805.00,267.58)(1.062,0.492){19}{\rule{0.936pt}{0.118pt}}
\multiput(805.00,266.17)(21.057,11.000){2}{\rule{0.468pt}{0.400pt}}
\multiput(828.00,278.58)(0.970,0.492){21}{\rule{0.867pt}{0.119pt}}
\multiput(828.00,277.17)(21.201,12.000){2}{\rule{0.433pt}{0.400pt}}
\multiput(851.00,290.58)(0.893,0.493){23}{\rule{0.808pt}{0.119pt}}
\multiput(851.00,289.17)(21.324,13.000){2}{\rule{0.404pt}{0.400pt}}
\multiput(874.00,303.58)(0.853,0.493){23}{\rule{0.777pt}{0.119pt}}
\multiput(874.00,302.17)(20.387,13.000){2}{\rule{0.388pt}{0.400pt}}
\multiput(896.00,316.58)(0.827,0.494){25}{\rule{0.757pt}{0.119pt}}
\multiput(896.00,315.17)(21.429,14.000){2}{\rule{0.379pt}{0.400pt}}
\multiput(919.00,330.58)(0.771,0.494){27}{\rule{0.713pt}{0.119pt}}
\multiput(919.00,329.17)(21.519,15.000){2}{\rule{0.357pt}{0.400pt}}
\multiput(942.00,345.58)(0.771,0.494){27}{\rule{0.713pt}{0.119pt}}
\multiput(942.00,344.17)(21.519,15.000){2}{\rule{0.357pt}{0.400pt}}
\multiput(965.00,360.58)(0.678,0.495){31}{\rule{0.641pt}{0.119pt}}
\multiput(965.00,359.17)(21.669,17.000){2}{\rule{0.321pt}{0.400pt}}
\multiput(988.00,377.58)(0.648,0.495){31}{\rule{0.618pt}{0.119pt}}
\multiput(988.00,376.17)(20.718,17.000){2}{\rule{0.309pt}{0.400pt}}
\multiput(1010.00,394.58)(0.639,0.495){33}{\rule{0.611pt}{0.119pt}}
\multiput(1010.00,393.17)(21.732,18.000){2}{\rule{0.306pt}{0.400pt}}
\multiput(1033.00,412.58)(0.605,0.495){35}{\rule{0.584pt}{0.119pt}}
\multiput(1033.00,411.17)(21.787,19.000){2}{\rule{0.292pt}{0.400pt}}
\multiput(1056.00,431.58)(0.546,0.496){39}{\rule{0.538pt}{0.119pt}}
\multiput(1056.00,430.17)(21.883,21.000){2}{\rule{0.269pt}{0.400pt}}
\multiput(1079.00,452.58)(0.546,0.496){39}{\rule{0.538pt}{0.119pt}}
\multiput(1079.00,451.17)(21.883,21.000){2}{\rule{0.269pt}{0.400pt}}
\multiput(1102.58,473.00)(0.496,0.521){41}{\rule{0.120pt}{0.518pt}}
\multiput(1101.17,473.00)(22.000,21.924){2}{\rule{0.400pt}{0.259pt}}
\multiput(1124.00,496.58)(0.498,0.496){43}{\rule{0.500pt}{0.120pt}}
\multiput(1124.00,495.17)(21.962,23.000){2}{\rule{0.250pt}{0.400pt}}
\multiput(1147.58,519.00)(0.496,0.542){43}{\rule{0.120pt}{0.535pt}}
\multiput(1146.17,519.00)(23.000,23.890){2}{\rule{0.400pt}{0.267pt}}
\multiput(1170.58,544.00)(0.496,0.564){43}{\rule{0.120pt}{0.552pt}}
\multiput(1169.17,544.00)(23.000,24.854){2}{\rule{0.400pt}{0.276pt}}
\multiput(1193.58,570.00)(0.496,0.587){43}{\rule{0.120pt}{0.570pt}}
\multiput(1192.17,570.00)(23.000,25.818){2}{\rule{0.400pt}{0.285pt}}
\multiput(1216.58,597.00)(0.496,0.660){41}{\rule{0.120pt}{0.627pt}}
\multiput(1215.17,597.00)(22.000,27.698){2}{\rule{0.400pt}{0.314pt}}
\multiput(1238.58,626.00)(0.496,0.653){43}{\rule{0.120pt}{0.622pt}}
\multiput(1237.17,626.00)(23.000,28.710){2}{\rule{0.400pt}{0.311pt}}
\multiput(1261.58,656.00)(0.496,0.697){43}{\rule{0.120pt}{0.657pt}}
\multiput(1260.17,656.00)(23.000,30.637){2}{\rule{0.400pt}{0.328pt}}
\multiput(1284.58,688.00)(0.496,0.719){43}{\rule{0.120pt}{0.674pt}}
\multiput(1283.17,688.00)(23.000,31.601){2}{\rule{0.400pt}{0.337pt}}
\multiput(1307.58,721.00)(0.496,0.763){43}{\rule{0.120pt}{0.709pt}}
\multiput(1306.17,721.00)(23.000,33.529){2}{\rule{0.400pt}{0.354pt}}
\multiput(1330.58,756.00)(0.496,0.822){41}{\rule{0.120pt}{0.755pt}}
\multiput(1329.17,756.00)(22.000,34.434){2}{\rule{0.400pt}{0.377pt}}
\multiput(1352.58,792.00)(0.496,0.830){43}{\rule{0.120pt}{0.761pt}}
\multiput(1351.17,792.00)(23.000,36.421){2}{\rule{0.400pt}{0.380pt}}
\multiput(1375.58,830.00)(0.496,0.874){43}{\rule{0.120pt}{0.796pt}}
\multiput(1374.17,830.00)(23.000,38.349){2}{\rule{0.400pt}{0.398pt}}
\put(281,166){\raisebox{-.8pt}{\makebox(0,0){$\Diamond$}}}
\put(548,197){\raisebox{-.8pt}{\makebox(0,0){$\Diamond$}}}
\put(671,207){\raisebox{-.8pt}{\makebox(0,0){$\Diamond$}}}
\put(851,240){\raisebox{-.8pt}{\makebox(0,0){$\Diamond$}}}
\put(725,230){\raisebox{-.8pt}{\makebox(0,0){$\Diamond$}}}
\put(957,357){\raisebox{-.8pt}{\makebox(0,0){$\Diamond$}}}
\put(1178,533){\raisebox{-.8pt}{\makebox(0,0){$\Diamond$}}}
\put(1393,849){\raisebox{-.8pt}{\makebox(0,0){$\Diamond$}}}
\end{picture}

%% file: hep.bbl
\newpage

\begin{thebibliography}{99}
\bibitem{SVZ} M. A. Shifman, A. I. Vainshtein and V. I. Zakharov, Nucl.
Phys. B {\bf 147}, 385, 448 (1979).
\bibitem{shif} M. A. Shifman in {\it Lectures and reprint series: 
``Vacuum Structure and QCD Sum Rules"}, ed. M. A. Shifman, North Holland, 1992.
\bibitem{rry} L. J. Reinders H. R. Rubinstein and S. Yazaki, Phys. Rep. 
 {\bf 127}, 1 (1985). 
\bibitem{dd} M. Dey and J. Dey, {\it Nuclear and Particle Physics: 
The Changing Interface}, Chapter 9, Springer Verlag, 1993.
\bibitem{ioff81} B. L.Ioffe, Nucl. Phys.  B 188 (1981) 591, 191(1981) 591 (E);
B. L. Ioffe, Z. Phys.  C {\bf 18}. 67 (1983).
\bibitem{smil} B. L. Ioffe and A. V. Smilga,  Phys. Lett.  B {\bf 133}, 436
(1983); B. L. Ioffe and A. V. Smilga,  Nucl. Phys. B {\bf 216}, 373 (1983).
\bibitem{mn} M. Nielsen, L. A. Barreiro, C. O. Escobar and R. Rosenfeld,
Phys. Rev. D {\bf 53}, 3620 (1996).
\bibitem{re} L. J. Reinders, H. R. Rubinstein and S. Yazaki , Phys. Lett. B
{\bf 120}, 209 (1983); {\bf 122}, 487(E) (1983).
\bibitem{bi} V. M. Belyaev and B. L. Ioffe, Sov. Phys. JETP {\bf 56} 493
(1982).
\bibitem{jin1} X. Jin, M. Nielsen and J. Pasupathy, Phys. Rev. {\bf D 51}
3688 (1995).
\bibitem{jin2} X. Jin  and J. Tang, Chirality and Reliability of Baryon QCD
Sum Rules, preprint MIT-CTP-2601, hep-ph/9701230, submitted to Phys. Rev. D. 
\bibitem{bel} V. M. Belyaev and B. L. Ioffe, Sov. Phys. JETP {\bf 57} 716
(1983).
\bibitem{smilga} A. V. Smilga in {\it Lectures and reprint series: 
``Vacuum Structure and QCD Sum Rules"}, ed. M. A. Shifman, North Holland,
1992, p. 492.
\bibitem{teren}M. V. Terent'ev, Sov. J. Nucl. Phys.  {\bf 24}, 106 (1976); 
L.A. Kondratyuk and M.V.Terent'ev, Sov. J.  Nucl. Phys.  {\bf 31}, 561 (1980).
\bibitem{keispol}B.D.Keister and W.N. Polizou, Adv. Nucl. Phys.  {\bf 20}, 225
(1991).
\bibitem{wil90}R.J.Perry, A.Harindranath and K.G. Wilson, Phys. Rev. Lett.
 {\bf 65}, 2959 (1990).
\bibitem{pach95} W.R.B. de Araujo, J.P.B.C. de Melo and T.Frederico, Phys.
Rev.  C {\bf 52}, 2733 (1995).
\bibitem{fm} T. Frederico and G. A. Miller, Phys. Rev.  D {\bf 50}, 210 
(1994). 
\bibitem{wein} S. Weinberg, Phys. Rev.  {\bf 150}, 1313 (1966).
\bibitem{tob92} T. Frederico, Phys. Lett.  B {\bf 282}, 409 (1992); 
S. K. Adhikari, L. Tomio and  T.Frederico, Ann. Phys. {\bf 235}, 77 (1994).
\bibitem{fadd} L. D. Faddeev, Sov. Phys. JETP  {\bf 12}, 1014 (1960).
\bibitem{renor}  S.K. Adhikari, T. Frederico and I.D. Goldman, Phys. Rev. 
Lett. {\bf 74}, 487 (1995).
\bibitem{ban94} M. K. Banerjee, Prog. Part. Nucl. Phys. {\bf 31} 77 (1993).
\bibitem{shur} E. V. Shuryak and J. L. Rosner, Phys. Lett.  B {\bf 218}, 72 
(1989). \  M. Dey and J. Dey, {\it Nuclear and Particle Physics: The Changing
Interface}, Chapter 6, Springer Verlag, 1993.
\bibitem{dor} A. E. Dorokov and N. I. Kochelev, Z. Phys.  C {\bf 46}, 281
(1990).
\end{thebibliography}
